\documentclass[aps,prl,twocolumn,superscriptaddress,showpacs]{revtex4}
\usepackage{amssymb}
\usepackage{amsmath}
\usepackage{graphicx}
\usepackage{natbib}
\begin{document}

\title{Nonadiabatic quantum control of a semiconductor charge qubit}
\author{Y. Dovzhenko}
\author{J. Stehlik}
\author{K. D. Petersson}
\author{J. R. Petta}
\affiliation{Department of Physics, Princeton University, Princeton, NJ 08544, USA}
\author{H. Lu}
\author{A. C. Gossard}
 \affiliation{Materials Department, University of California at Santa Barbara, Santa Barbara, CA 93106, USA}

\date{\today}

\begin{abstract}
We demonstrate multipulse quantum control of a single electron charge qubit. The qubit is manipulated by applying nonadiabatic voltage pulses to a surface depletion gate and readout is achieved using a quantum point-contact charge sensor. We observe Ramsey fringes in the excited-state occupation in response to a $\pi/2$ -- $\pi/2$ pulse sequence and extract $T_2^*\sim60~$ps away from the charge degeneracy point. Simulations suggest these results may be extended to implement a charge echo by reducing the interdot tunnel coupling and pulse rise time, thereby increasing the nonadiabaticity of the pulses.
\end{abstract}

\pacs{85.35.Gv, 03.67.Lx, 73.21.La}

\maketitle
The goal of quantum computation has inspired a number of recent experiments in solid-state physics. A few promising qubit candidates created so far include quantum dot charge and spin qubits, as well as a variety of superconducting devices \cite{Hayashi03,Petta05,Nakamura99,Vion02,Martinis02,Wallraff04}. One significant challenge facing all of these architectures is the loss of coherence due to interactions with environmental degrees of freedom. Since the early days of nuclear magnetic resonance, physicists have developed pulse sequences for mitigating environmental interactions \cite{Hahn50}. In the context of quantum information science, the concept has been expanded in the form of dynamic decoupling pulse sequences \cite{Viola99,Uhrig07}. Dynamic decoupling has been employed to extend coherence in a wide range of systems, including nitrogen vacancy (NV) centers in diamond, trapped ions, and electron spins in quantum dots \cite{Biercuk09,Lange10,Barthel10,Bluhm11}.

The efficiency with which dynamic decoupling maintains coherence depends on the specific pulse sequence used as well as the spectral characteristics of the noisy environment \cite{Biercuk09}. A comprehensive study of charge noise has been performed in superconducting Cooper-pair-box charge qubits \cite{Nakamura02,Lehnert03,Ithier05}. These experiments have shown Larmor precession of the qubit state, probed inhomogeneous dephasing through Ramsey spectroscopy, and demonstrated charge echo. Furthermore, charge noise at different frequency scales was studied by analyzing the amplitude of the echo signal as a function of the evolution time. In semiconductors, coherence of a charge qubit has been demonstrated using single non-adiabatic pulses in the many-electron and one-electron regimes \cite{Hayashi03,Fujisawa04,Petta04,Petersson10}. However, in order to arrive at a precise expression for the noise spectral density, elaborate pulse sequences building on Ramsey spectroscopy and echo are needed \cite{Bylander11}. In this Rapid Communication we demonstrate multipulse control of a charge qubit, paving the way for dynamic decoupling.

\begin{figure}[t]
\begin{center}
\includegraphics[scale=1]{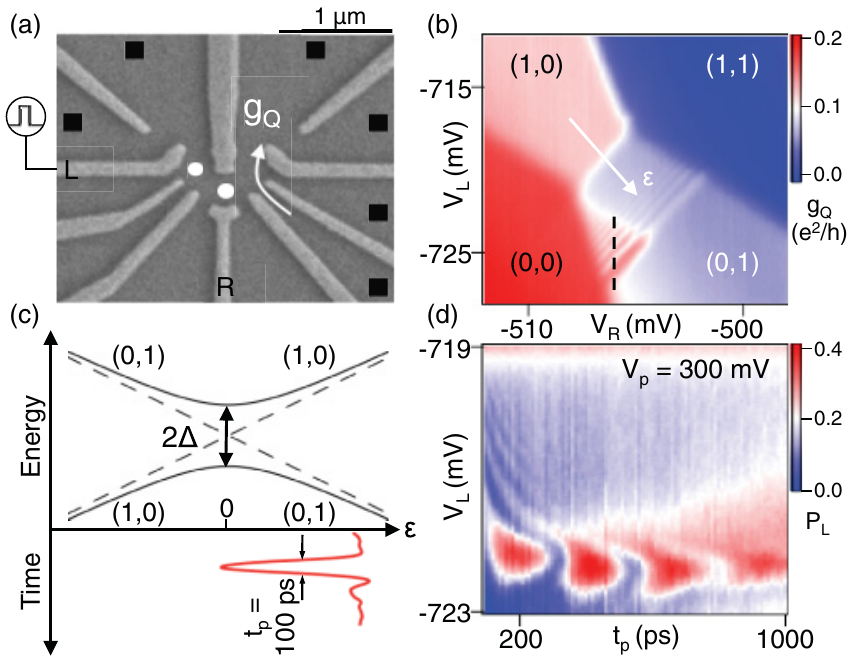}
\caption{\label{fig1} (a) Scanning electron microscope image of a sample similar to the one measured. The triple quantum dot is configured as a double dot, with the third dot operated as a charge sensor. Ohmic contacts to the 2DEG are marked by black squares. Voltage pulses are applied to the left gate (labeled L). (b) Charge sensor conductance, g$_Q$, measured near the (0,1) -- (1,0) charge transition with nonadiabatic pulses applied. The dashed line shows a typical range of gate voltages where qubit state readout is performed. (c) Qubit energy-level diagram. A typical voltage pulse is shown in the lower panel. (d) Coherent oscillations are obtained by varying the pulse width $t_p$.}
\end{center}
\vspace{-1.1 cm}
\end{figure}

Figure 1(a) shows a scanning electron microscope image of a sample identical to the one measured. A two-dimensional electron gas (2DEG) with charge density 2 $\times$ 10$^{11}$/cm$^2$ is located 110 nm below the surface of the wafer. Surface depletion gates are arranged in a triple quantum dot geometry. For this experiment, a single electron is isolated in a double quantum dot and the third dot is used as a highly sensitive quantum point contact (QPC) charge detector. All measurements are performed in a dilution refrigerator with an electron temperature of 90 mK. The two basis states describing the qubit occupation are $|L\rangle=(1, 0)$ and $|R\rangle=(0,1)$, which correspond to the electron occupying the left or right dot. These states form an effective two-level system described by the Hamiltonian
\begin{equation}
H = \frac{1}{2}\varepsilon\sigma_z+\Delta\sigma_x,
\label{hamiltonian}
\end{equation}
where $\Delta$ is the tunnel coupling and $\varepsilon$ is the detuning. The tunnel coupling $\Delta$ results in an avoided crossing with a minimum energy splitting of $2\Delta$ at zero detuning. We adjust the energy difference between $|L\rangle$ and $|R\rangle$ by sweeping the gate voltages along the detuning axis [see the white arrow in Fig.\ 1(b)]. However, for quantum control we only drive the left gate to avoid the requirement of generating synchronized $\sim$ 100 ps pulses. The resulting conversion between gate voltage and detuning is $\varepsilon \propto \alpha V_L$, where $\alpha$ = -54 $\mu\mathrm{eV/mV}$.

We first demonstrate coherent control of the qubit by driving it with a single pulse \cite{Hayashi03,Petersson10}. The qubit is initialized at positive detuning in $|R\rangle$, a charge eigenstate. A nonadiabatic voltage pulse of amplitude $V_p$ and width $t_p$ (measured at the output of the pulse generator) drives the system to $\varepsilon = 0$. At $\varepsilon = 0$ the eigenstates are $(|L\rangle\pm|R\rangle)/\sqrt{2}$, and the qubit evolves according to a $\sigma_x$ rotation on the Bloch sphere. After the free evolution time $t_p$, the final charge state is a superposition of $|L\rangle$ and $|R\rangle$, with the relative weighting set by $V_p$ and $t_p$. An example of a pulse used in this part of the experiment is shown in Fig.\ 1(c). Figure 1(b) displays the charge sensor conductance, $g_Q$, as a function of the dc voltages, $V_L$ ($V_R$) on the left (right) gate electrodes. Pulse parameters for this data set are $t_p=130$ ps and $V_p$ = 700 mV. Several bright lines at positive detuning indicate conditions for which the qubit precesses to the excited state, $|L\rangle$, during the pulse.

We observe coherent charge oscillations by continuously measuring the charge sensor conductance, $g_Q$, as a function of $V_L$ and $t_p$, as shown in Fig.\ 1(d). The oscillations are robust when the qubit is driven to the charge degeneracy point, consistent with previous reports \cite{Vion02,Fujisawa04,Petersson10}. Readout is performed along the dashed line in Fig.\ 1(b), where the qubit can be trapped in the (0, 0) state during readout, leading to enhanced visibility of the oscillations \cite{Petersson10}. The excited state probability, $P_L$, is determined by normalizing the charge sensor conductance to the (0,0) and (1,0) plateaus in the charge stability diagram. We choose the length of each pulse sequence, including the measurement interval, to equal $\sim$ 30 ns. We extract a tunnel coupling $\Delta/h\sim2$ GHz from the coherent oscillations in Fig.\ 1(d).

\begin{figure}[t]
\begin{center}
\includegraphics[scale=1]{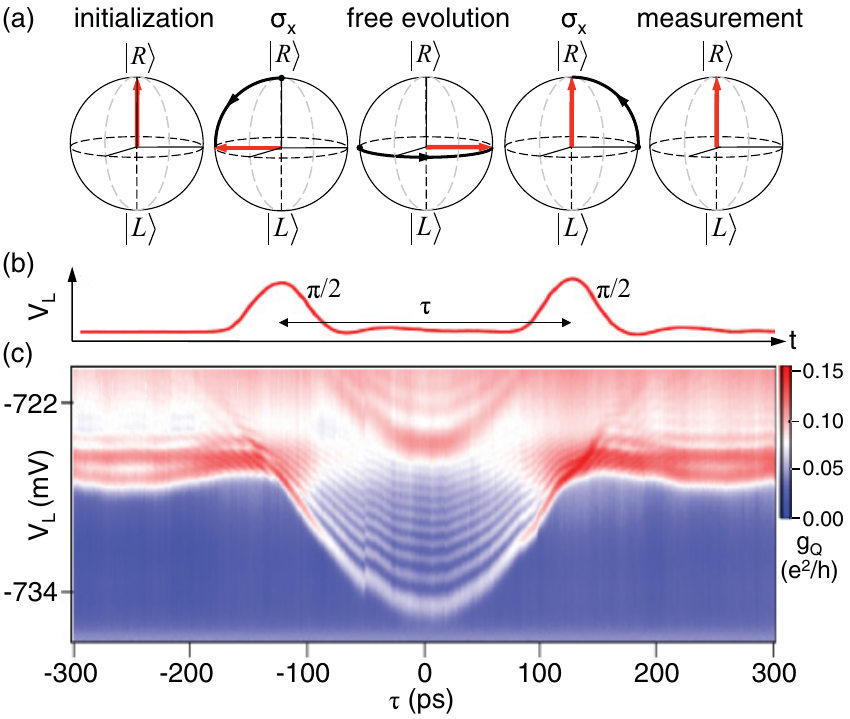}
\caption{\label{fig2} (a) Ramsey fringe pulse sequence depicted in the Bloch sphere representation. The qubit is initialized in $|R\rangle$. A $\sigma_x = \pi/2$ pulse rotates the Bloch vector to the equator. The qubit then freely evolves for a variable time $\tau$, resulting in a $\sigma_z$ rotation on the Bloch sphere. The final state is determined using charge sensing after a second $\sigma_x = \pi/2$ pulse is applied. (b) A typical voltage pulse profile with $t_{\pi/2} = 130$ ps. (c) QPC conductance measured as a function of delay, $\tau$, for a pulse amplitude $V_p$ = 900 mV. The two pulses partially overlap for $|\tau|$ $<$ 140 ps, resulting in Larmor precession similar the data in Fig.\ 1(d). Ramsey fringes are observed for longer delay times.}
\end{center}
\vspace{-1.0 cm}
\end{figure}

In the single pulse measurement, Larmor precession occurs near the charge degeneracy point, where the level splitting is first-order insensitive to charge noise. In order to probe decoherence at arbitrary values of the detuning, we use Ramsey interference \cite{Vion02}. We separate two $\sigma_x = \pi/2$ rotations by a free evolution period of variable length $\tau$ to form the Ramsey pulse sequence. A typical voltage pulse profile is shown in Fig.\ 2(b), excluding the $\sim$ 30 ns measurement interval. This pulse sequence is generated by combining the two output channels of an Agilent 81134A pulse generator. The duration and amplitude of each pulse is selected to result in an approximate $\pi/2$ rotation at the ``sweet spot'' in the energy level diagram.

Evolution of the qubit state is depicted in the Bloch sphere represenation shown in Fig.\ 2(a). At the beginning of each pulse sequence the qubit is prepared at positive detuning in $|R\rangle$. The first $\sigma_x= \pi/2$ pulse, with width $t_{\pi/2}$, rotates the Bloch vector into the $x$-$y$ plane. Following the first pulse, the Bloch vector freely precesses about the $z$ axis with frequency $f=\sqrt{\varepsilon^2+4\Delta^2}/h$. This free evolution is interrupted by a second $\sigma_x = \pi/2$ rotation. In the absence of any $\sigma_z$ rotation, the two back-to-back $\sigma_x$ rotations convert $|R\rangle$ to $|L\rangle$. Adding a delay between the $\sigma_x$ rotations results in oscillations in the excited state probability known as Ramsey fringes.

Measurements of the charge sensor conductance exhibit coherent oscillations as a function of $V_L$ and $\tau$, as shown in Fig.\ 2(c). For this data set $t_{\pi/2}$ = 122 ps and the delay ranges from -300 ps to +300 ps. The two $\sigma_x = \pi/2$ pulses overlap for $|\tau|$ $\sim$ 0, resulting in a pulse that is approximately twice the height of a single pulse. As a result, the oscillations in the vicinity of $\tau$ = 0 are a signature of Larmor precession, similar to the data in Fig.\ 1(b). For $|\tau|$ $>$ 140 ps the two $\sigma_x$ pulses are separated by a free evolution period, resulting in Ramsey fringes that decay with increasing delay.

To demonstrate the reproducibility of the experiment, we show a second Ramsey data set in Fig.\ 3(c), with pulse parameters $t_{\pi/2}$ = 130 ps and $V_p$ = 700 mV. In order to quantitatively analyze the data we extract the excited-state probability, $P_L$, as a function of $\tau$ along the dashed line in Fig.\ 3(c). We subtract a slowly varying background from these data and show the resulting Ramsey fringes in Fig.\ 3(a). We fit the oscillations to a cosine function with an amplitude that exhibits a Gaussian decay, consistent with a simple $1/f$ noise model \cite{Petersson10},

\begin{equation}
\Delta P_L=A_0 \exp\left[-\left(\frac{\tau-\tau_0}{\mathrm{T}_2^*}\right)^2\right] \cos\left[2\pi f\left(\tau-\tau_0\right)\right].
\label{dephasing}
\end{equation}
Here $A_0$ is the oscillation amplitude, $T_2^*$ is the inhomogeneous dephasing time, $\tau_0$ = 135 ps is a fixed offset time, and $f$ is the oscillation frequency used in the fitting. For the data shown in Fig.\ 3(a) we extract $T_2^*$ = 60 $\pm$ 10 ps. Similar values of $T_2^*$ were obtained from several other cuts at different values of $V_L$. Figure 3(b) shows the extracted Ramsey frequencies as a function of $V_L$ along with a theoretical prediction based on the interdot tunnel coupling $\Delta/h$ = 2 GHz and lever arm conversion between gate voltage and energy.

The evolution of the charge qubit is modeled by numerically integrating the time-dependent Schrodinger equation using voltage pulse profiles acquired at the output of the pulse generator, tunnel coupling measured from the single pulse Larmor precession data set, and the Hamiltonian in Eq.\ \ref{hamiltonian}. Figure 3(d) shows the calculated excited state probability as a function of $V_L$ and $\tau$. In order to account for charge noise, each vertical cut through the simulation is convolved with a Gaussian $\frac{1}{\sqrt{2\pi \sigma_\varepsilon^2}}e^{-\frac{(\alpha V_L)^2}{2\sigma_\varepsilon^2}}$ with $\sigma_\varepsilon = 7.3~\mu\mathrm{eV}$. The value of $\sigma_\varepsilon$ is chosen such that the Ramsey fringes at the ``sweet spot" have equal values of $T_2^*$ in both the data and the simulation. While the characteristic decay times are similar, the amplitude of the oscillations in the simulation is larger because charge relaxation is not taken into account. The value of $\sigma_\varepsilon$ extracted from the simulation suggests a dephasing time of $\sim$ 130 ps according to the relation $T_2^*\approx\frac{\sqrt{2}\hbar}{\sigma_\varepsilon}$, \cite{Petersson10} which is greater than the value of 60 ps obtained from the data. It is possible that 60 ps underestimates the true $T_2^*$ because additional dephasing during the $\pi/2$ pulses is not taken into account in Eq.\ \ref{dephasing}. For comparison, setting $\tau_0 = 0$ in Eq.\ \ref{dephasing} leads to T$_2^*$ = 130 ns \cite{Nakamura02}.

\begin{figure}[t]
\begin{center}
\includegraphics[scale=1]{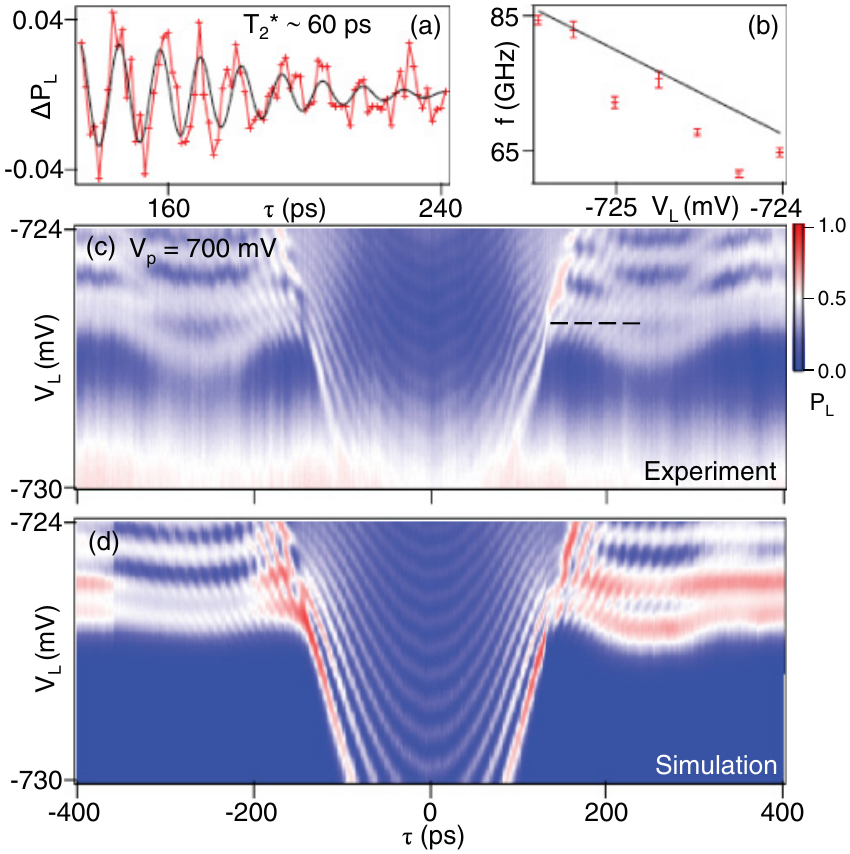}
\caption{\label{fig3} (a) Ramsey fringes with a fit to Eq.\ \ref{dephasing}. The value for $T_2^*$ extracted from the fit is 60 $\pm$ 10 ps. (b) Ramsey frequency as a function of $V_L$ extracted from experimental data (in red). The black line shows the expected frequency based on spectroscopy of the energy-level diagram. (c) Excited-state occupation $P_L$ as a function of $V_L$ and $\tau$. The dashed line shows the location of the cut in (a). (d) Numerical simulation of the excited-state occupation generated using pulse profiles acquired at the output of the pulse generator, plotted on the same color scale as the experimental data in (c).}
\end{center}
\vspace{-1.0 cm}
\end{figure}

Coherence during a Ramsey experiment is limited by inhomogeneities in the precession frequency due to low-frequency charge noise. In this experiment each data point corresponds to an average over $\sim$ three million iterations of the pulse sequence. Since the precession frequency is a function of a noisy detuning parameter, each iteration of the pulse sequence results in a slightly different final state. In order to cancel out the effects of varying frequency, we apply a charge-echo pulse sequence similar to a NMR spin echo \cite{Hahn50}. Charge qubit evolution with the charge-echo pulse sequence is depicted in Fig.\ 4(a). A $\sigma_x= \pi/2$ pulse rotates the Bloch vector into the $x$-$y$ plane and then the Bloch vector freely precesses during a time period $\tau+\delta t$. We then apply a $\sigma_x = \pi$ pulse to reverse the evolution. This is followed by a second free evolution period of length $\tau-\delta t$. A final $\sigma_x= \pi/2$ pulse is applied before the excited state occupation is measured. Since the Bloch vector spends approximately equal amounts of time rotating in both directions, fluctuations in precession frequency now have a smaller impact on the dephasing time. The refocusing pulse is most effective when $\delta t$ = 0.

A voltage pulse profile used to implement the charge echo is illustrated in Fig.\ 4(b). For this data set we set $V_p = 900$ mV, $t_{\pi/2} = 122$ ps, $t_\pi = 190$ ps, and $\tau=900$ ps. Doubling the pulse width at the pulse generator does not necessarily double the evolution time at the sample due to the finite rise time of the pulses. Therefore the length of a $\pi$ pulse is not equal to twice the length of a $\pi/2$ pulse. Moreover, the finite rise time reduces the pulse amplitude at short pulse lengths.  We picked a relatively small initial value of $\tau$ because longer delay allows more inhomogeneity in evolution frequency within a single pulse sequence. The excited state probability is expected to show oscillations as a function of $\delta t$ around $\delta t = 0$ due to the charge echo. Experimental data do not display an echo, as illustrated in Fig.\ 4(c) and an expanded plot in Fig.\ 4(d)(i).

\begin{figure}[t]
\begin{center}
\includegraphics[scale=1]{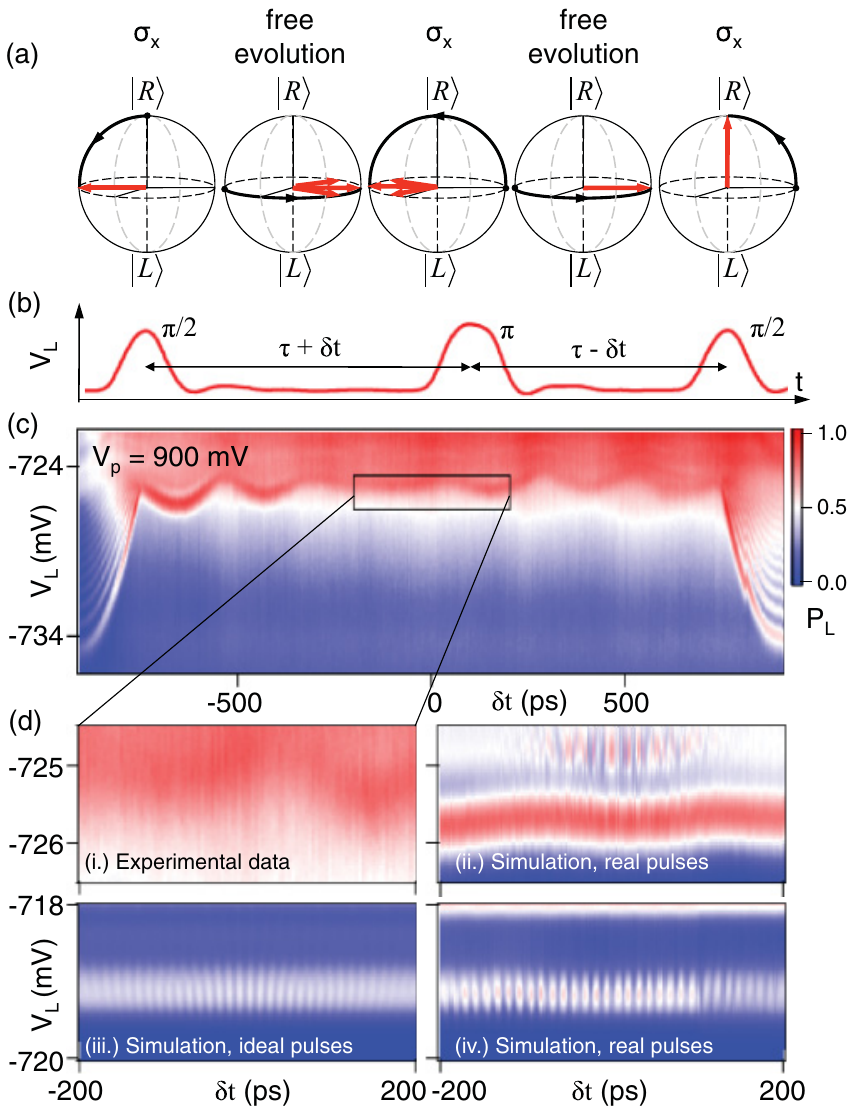}
\caption{\label{fig4} (a) Evolution of the charge qubit during an ideal charge-echo pulse sequence. (b) A typical pulse sequence, as acquired at the output of the pulse generator. (c) Excited state probability, $P_L$, as a function of $V_L$ and the displacement $\delta t$. (d) Expanded plots of $P_L$ as a function of $V_L$ and $\delta t$ near $\delta t$ = 0 for: (i) Experimental data (c), (ii) Simulation using experimental pulses, (iii) Simulation with $\Delta/h = 0.5$ GHz and using ideal Gaussian pulses (see text) with $V_p$ =  300 mV and a rise time of 60 ps, (iv) Simulation with $\Delta/h = 0.5$ GHz using experimental pulses with $V_p$ = 300 mV.}
\end{center}
\vspace{-1.0 cm}
\end{figure}

We simulated the qubit evolution using voltage pulse profiles acquired at the output of the pulse generator in order to understand the absence of a charge echo. The simulations in Fig.\ 4(d)(ii) suggest the presence of charge oscillations with a period $\delta t$ of approximately 4 ps, which are too closely spaced to be observed in this sample due to QPC charge noise. To further investigate the oscillation visibility, we simulated qubit evolution under the influence of perfect square pulses that are convolved with a Gaussian to give a 60-ps, 20$\%$ -- 80 $\%$ rise time (ideal pulses). For sufficiently small tunnel coupling, $\Delta/h = 0.5$ GHz, the simulations predict an echo signal, as shown in Fig.\ 4(d)(iii). Based on this observation, we simulated qubit evolution for experimental pulses with $V_p$ = 300 mV and $\Delta/h = 0.5$ GHz. These simulations are shown in Fig.\ 4(d)(iv) and indicate that a charge echo may be observable with sufficiently fast pulse rise times.

The dynamic decoupling pulse sequences developed by Uhrig assume hard pulses are applied to the sample \cite{Uhrig07}. For charge qubits, the time scales for quantum evolution impose very strict technical requirements on pulse generation capability. First, the voltage pulse must be nonadiabatic, otherwise the qubit would simply remain in the ground state. In this case, the minimum rise time is limited to $\sim$ 60 ps by the pulse generator. Second, for a fixed detuning, the frequency of the echo signal occurs at twice the Larmor precession frequency, since the relevant time difference is 2$\delta t$ \cite{Nakamura02}. In order to observe a clear echo, the oscillations must be spaced by a time interval large enough to allow the oscillations to remain visible in the presence of QPC charge noise. For comparison, a charge echo has been observed at $\sim$ 100 GHz in a superconducting qubit \cite{Nakamura02}. Lastly, the entire pulse sequence must occur within a timescale set by $T_2$ $<$ 10 ns \cite{Hayashi03,Petersson10}. All of this suggests: 1) small tunnel couplings, but large enough to observe several free precessions within $T_2$, 2) small pulse amplitudes to reduce the echo oscillation frequency, and 3) extremely fast rise times, as to maintain sufficient level velocity for high visibility coherent oscillations. Further technical improvements will be required to bring quantum control of the semiconductor charge qubit to the level attained in spin systems, where the temporal dynamics are much slower.

\begin{acknowledgments}
Research at Princeton is supported by the Sloan and Packard Foundations, the NSF through the PCCM (DMR-0819860) and CAREER award (DMR-0846341), and DARPA QuEST (HR0011-09-1-0007). Work at UCSB was supported by DARPA (N66001-09-1-2020) and the UCSB NSF DMR MRSEC.
\end{acknowledgments}



\end{document}